\documentclass[sigconf]{acmart}

\AtBeginDocument{%
  \providecommand\BibTeX{{%
    \normalfont B\kern-0.5em{\scshape i\kern-0.25em b}\kern-0.8em\TeX}}}

\usepackage{pgfplots} 

\usepackage{hyperref}
\usepackage{enumitem}
\usepackage{stfloats}
\usepackage{threeparttable}

\usepackage{xcolor}
\usepackage{url}
\usepackage{soul}
\usepackage{colortbl}
\usepackage{adjustbox}
\usepackage{subfigure}
\usepackage{caption}
\usepackage{subcaption}
\usepackage{xcolor}
\usepackage{refcount}
\usepackage{float}
\usepackage{siunitx}
\usepackage{booktabs}
\usepackage{graphicx}
\usepackage{multirow}
\usepackage{makecell} 
\usepackage[most]{tcolorbox} 
\usetikzlibrary{tikzmark,calc}
\usepackage{calc}

\definecolor{vlightgray}{gray}{0.9}

\DeclareRobustCommand*{\IEEEauthorrefmark}[1]{%
  \raisebox{0pt}[0pt][0pt]{\textsuperscript{\footnotesize #1}}%
}

\hypersetup{
 colorlinks=true,   
 citecolor=black,
 urlcolor=blue,
 linkcolor=black,
 bookmarksnumbered=true,     
 bookmarksopen=true,         
 bookmarksopenlevel=1,          
 pdfstartview=Fit,           
 pdfpagemode=UseOutlines,    
 pdfpagelayout=TwoPageRight 
}

\newcolumntype{P}[1]{>{\raggedright\arraybackslash}p{#1}}

\newcommand{\pQuote}[2]{{\color{TekheletPurple}\textit{``#1''}~({#2})}}

\newif\ifdraft
\draftfalse 

\newif\ifrevising
   \revisingfalse 

 \newcommand{\deleted}[1]{{\ifrevising{\relax}\else\relax\fi}}

\definecolor{TekheletPurple}{HTML}{572F96}
\newtcolorbox{takeawayBox}{
  breakable,
  enhanced,
  sharp corners,
  colback=TekheletPurple!20!white,             
  colframe=TekheletPurple!20!white,            
  boxrule=0.5pt,
  left=3pt,
  right=3pt,
  top=3pt,
  bottom=3pt,
  before skip=1pt,
  after skip=1pt,
  fonttitle=\bfseries\small,
  coltitle=black,
  attach title to upper={},
  separator sign={\ },
}

\makeatletter
\providecommand\nopunct{\@addpunct{}}
\makeatother

\copyrightyear{2026}
\acmYear{2026}
\setcopyright{cc}
\setcctype{by}
\acmConference[ICSE-SEIP '26]{2026 IEEE/ACM 48th International Conference on Software Engineering}{April 12--18, 2026}{Rio de Janeiro, Brazil}

\begin{document}

\title{Understanding Developers' Attitudes Towards Daily Tasks, AI Autonomy, and Preferred Human-AI Interaction Aspects}
\title{AI Where It Matters: \\Human-Centric AI for Software Engineering}
\title{Human-Centric AI for Software Engineering: A Study of Developer Task Appraisals, AI Autonomy \& Responsible Design}
\title{Human-Centric AI for Software Engineering: Task Appraisals, \\AI Preferences \& Responsible Design Priorities}
\title{AI Where It Matters: \\Where, Why, and How Developers Want AI Support in Daily Work}

\affiliation{Rudrajit Choudhuri\IEEEauthorrefmark{1}, \hspace{0.2em}  Carmen Badea\IEEEauthorrefmark{2}, \hspace{0.2em} Christian Bird\IEEEauthorrefmark{2}, \hspace{0.2em} Jenna L. Butler\IEEEauthorrefmark{2}, \hspace{0.2em} Robert DeLine\IEEEauthorrefmark{2}, \hspace{0.2em}Brian Houck\IEEEauthorrefmark{2}  
\country{}
}


\affiliation{
\institution{ \normalsize \IEEEauthorrefmark{1}Oregon State University, OR, USA. Email: choudhru@oregonstate.edu}
\country{}
}

\affiliation{
\institution{ \normalsize \IEEEauthorrefmark{2}Microsoft, WA, USA. Email: {cabadea, cbird, jennbu, rdeline, bhouck}@microsoft.com}
\country{}
}



%
\renewcommand{\shortauthors}{Choudhuri et al.}

\newcommand{\explaintwo}[1]{%
\par%
\noindent\fbox{%
    \parbox{\dimexpr\linewidth-2\fboxsep-2\fboxrule}{#1}%
}%
}

\settopmatter{printacmref=false}


\newcommand{\blue}[1]{\textcolor{blue}{#1}}



\begin{abstract}

Generative AI is reshaping software work, yet we lack clear guidance on where developers most need support and how to design it responsibly.
We report a large-scale, mixed-methods study of \textbf{N=860} developers examining where, why, and how they seek or limit AI help across SE tasks.
Using cognitive appraisal theory, we provide the first empirically validated mapping of developers' task appraisals to AI adoption patterns and Responsible AI (RAI) priorities. Appraisals predict AI openness and use, revealing distinct patterns: strong current use and demand for improvement in \textit{core work} (e.g., coding, testing); high demand to reduce \textit{toil} (e.g., documentation, operations); and clear limits for \textit{identity-} and \textit{relationship-centric work} (e.g., mentoring). RAI priorities vary by context: reliability and security for systems-facing tasks; transparency, alignment, and steerability to maintain control; and fairness and inclusiveness for human-facing work.
Our results offer concrete, contextual guidance for delivering AI where it matters to developers and their work.

\end{abstract}

%

\maketitle

\section{Introduction}
\label{sec:intro}

Developers increasingly work with generative AI tools (e.g., Copilot, Cursor) that promise faster delivery and reduced cognitive load~\cite{stackoverflow2024,bird2022taking,Vaithilingam2022Expectation,mckinsey2024}. Yet AI adoption in software engineering (SE) reveals a persistent tension: while capabilities are advancing rapidly~\cite{hou2024large}, AI integration often proceeds without clarity on \textit{where} developers need help, where they prefer retaining control, and \textit{how} to design responsible AI support~\cite{russo2024navigating, choudhuri2025guides, bird2022taking}. Without this clarity, we risk optimizing the wrong aspects of SE work. Industry evidence highlights this concern: developers using AI report higher satisfaction and frequent experiences of \textit{“flow”}, yet spend less time on work they consider valuable~\cite{dora2024}, potentially weakening professional identity and the quality judgments that define effective SE work.

In this paper, ``AI'' refers to developer tools powered by generative AI models, including commercial offerings like GitHub Copilot, Claude, Cursor, and bespoke in-house solutions that assist with various aspects of software development. 

Existing research has made important progress in understanding AI adoption in SE~\cite{russo2024navigating, khemka2024toward, choudhuri2025needs, pereira2025exploring}. Studies have documented task-level preferences~\cite{khemka2024toward, pereira2025exploring}, showing that workflow fit can outweigh perceived usefulness in early adoption decisions~\cite{russo2024navigating}. Other work has identified gaps between developers’ ideal and actual workweeks, highlighting toil-heavy activities as prime candidates for AI support~\cite{kumar2025time}. However, while these studies clarify \textit{which} tasks developers want automated, they do not explain \textit{why} or \textit{how}.

We argue that meaningful AI integration requires understanding how developers cognitively appraise different aspects of their work. Drawing on cognitive appraisal and work design theories~\cite{lazarus1991emotion, roseman2001appraisal, humphrey2007integrating}, we examine how developers' task appraisals along dimensions of relevance, identity congruence, accountability, and demands~\cite{Folkman1986Dynamics, humphrey2007integrating, lips2009discriminating} inform \textit{where}, \textit{why}, and \textit{how} they want AI to complement their workflows. Our investigation addresses:

\begin{itemize}
    \item \textbf{RQ1:} How do developers' task appraisals shape their openness to and use of AI tools? Where and why do they seek or limit AI support?
    
    \item \textbf{RQ2:} Which Responsible AI (RAI) design principles do developers prioritize for AI support in SE tasks, and how do these priorities vary with experience and AI dispositions?
    
\end{itemize}

We present findings from a large-scale mixed-methods study of \textbf{860} developers at Microsoft, showing how task appraisals predict AI adoption patterns (\textit{where/why}) and priorities for RAI design for AI support across SE tasks (\textit{how}). Using quantitative ratings across SE task categories, forced-choice prioritization of RAI principles, and thematic analysis of rationales, we map developers' AI needs/current usage patterns and the underlying psychological and professional considerations driving them.

More specifically, our findings reveal distinct task clusters that differ in their suitability for AI support. A quadrant map comparing support needs with current use exposes gaps between developer preferences and available tools, identifying opportunities for tool development. Further, we find that RAI priorities vary by context: for example, system-facing work requires stronger reliability and transparency than exploratory or creative work. These results yield an empirically grounded framework for calibrating AI assistance that preserves developer agency and sustains meaningful work.
\section{Related Work}
\label{sec: backg}

As AI tools enter development workflows, understanding what drives developers to adopt them has become a focal topic in SE research~\cite{russo2024navigating, bird2022taking, lambiase2025exploring, choudhuri2025guides}. Prior studies have applied technology-acceptance models (e.g., UTAUT~\cite{venkatesh2003user}) to understand AI adoption, finding that workflow compatibility and habitual use outweigh traditional factors such as performance or effort expectancy~\cite{russo2024navigating}. Trust also emerges as a key factor, shaped by both tooling capabilities and user dispositions (e.g., risk tolerance, technophilia)~\cite{choudhuri2025guides, johnson2023make, butler2025dear}.

More recently, studies have shifted from general adoption to investigating task-level differences~\cite{khemka2024toward, pereira2025exploring, kumar2025time}. \citet{lambiase2025exploring} found that AI receptivity was higher for artifact manipulation and information-retrieval tasks, but lower in collaborative contexts. In SE, \citet{pereira2025exploring} observed stronger AI adoption for code-intensive work with limited use in creative aspects. \citet{khemka2024toward} reported strong demand for AI in testing, debugging, documentation, and compliance. Complementing these findings, \citet{kumar2025time} compared developers' ideal versus actual time allocations, showing that toil-heavy activities (e.g., documentation, environment setup) correlated with reduced satisfaction and productivity. These tasks were disproportionately seen as work to minimize, positioning them as strong candidates for AI support.
Collectively, this literature indicates that AI adoption is not a monolith; it is calibrated to the nature of the task. Yet it stops short of probing the psychological rationales that shape delegability. For example, why does coding count as ``ideal'' time, while infrastructure work or rote refactoring does not?

Our study addresses this gap by shifting from solely a capability/fit perspective to a meaning-based account: developers ask not only \textit{“Can AI do this?”} but also \textit{“Should it?”} and \textit{“To what extent?”} We examine how developers cognitively appraise various aspects of their work to explain where, why, and how they seek or limit AI support (see \textsection \ref{want-AI}), revealing where human oversight remains essential even when AI is used.

Additionally, to our knowledge, this is the first study to examine developers' task-conditioned priorities for Responsible AI (RAI) design principles in AI-powered SE tools (see \textsection ~\ref{rq2-priorities}). We also report how these priorities vary by SE/AI experience and individuals' AI dispositions to guide adaptive, task- and user-sensitive design.


\section{Appraisal Foundations \& Hypotheses}
\label{sec:theory}

Individuals are meaning-makers; we actively seek significance and value in our experiences~\cite{lips2009discriminating}.
At work, we implicitly evaluate tasks by asking: \textit{Is this important to me? Does this align with what I want to do? Am I responsible if it fails? Can I handle its demands?} Cognitive appraisal theory~\cite{lazarus1991emotion, roseman2001appraisal} formalizes these judgments across dimensions of \textit{relevance/importance}, \textit{congruence with one’s motivations or identity}, \textit{accountability}, and \textit{cognitive demands}. These appraisals shape coping strategies~\cite{campbell2020cognitive} and predict downstream outcomes such as engagement, persistence, and discretionary effort~\cite{meyer1991three}. Complementing this, decades of work-design research~\cite{humphrey2007integrating, lips2009discriminating} show that job characteristics cluster into motivational (value, enjoyment), social (responsibility), and contextual (workload) factors, explaining substantial variance in work satisfaction and productivity~\cite{humphrey2007integrating}.

At this intersection, we focus on four appraisal drivers: \textbf{Value}, \textbf{Identity}, \textbf{Accountability}, and \textbf{Demands}. Value and Identity capture motivational aspects that make tasks meaningful~\cite{lips2009discriminating}; Accountability reflects the social stakes of responsibility~\cite{tetlock1983accountability}; and Demands index contextual difficulty and cognitive effort~\cite{bakker2007job}. These drivers shape how individuals perceive ownership, risk, and burden~\cite{koestner2002attaining, bailey2019review, bakker2007job}, thereby influencing whether and to what extent they seek support~\cite{shneiderman2020human, parasuraman2000model, lubars2019ask}. 
In SE, we hypothesize that developers’ openness to and use of AI are shaped by these drivers:

\textbf{Value} is the perceived significance of a task to project success, stakeholders, or personal goals~\cite{hackman1976motivation}, contributing to a belief that one's work matters~\cite{lips2022we, allan2019outcomes}. High-value tasks heighten attention, focus, and satisfaction, but also raise anxiety about failure~\cite{bailey2019review}. Such tasks attract tooling support, provided reliability is high~\cite{parasuraman2000model}. In SE, developers may welcome AI assistance to increase efficiency in valuable tasks, yet hesitate to cede control.

\textbf{H1}. \textit{Higher task value increases developers’ openness to AI support and usage.} We expect that developers seek AI support as a means to complement valuable tasks, rather than replacing them outright.

\textbf{Identity} alignment is the extent to which a task reflects one’s interests, expertise, or professional self-concept~\cite{hackman1976motivation, ryan2000self}. Such tasks are intrinsically motivating and foster a sense of purpose and ownership~\cite{koestner2002attaining, kahn1990psychological}, which can heighten reluctance to delegate them to AI~\cite{lubars2019ask}. Yet, identity can also increase engagement with tools that amplify one’s craft~\cite{shneiderman2020human}. Developers may resist ceding identity-defining work while strategically using AI to extend capabilities. 

\textbf{H2}. \textit{Higher task identity reduces developers’ openness to AI support, but can increase usage when AI complements expertise.}

\textbf{Accountability} refers to perceived responsibility and potential blame an individual feels for a task's outcome~\cite{tetlock1983accountability, lerner1999accounting}. High-accountability tasks carry serious reputational or organizational consequences (e.g., customer-facing failures).
Accountability Theory~\cite{tetlock1983accountability} suggests that when individuals anticipate (social) evaluation, they become more deliberate and information-seeking, turning to external aids as safeguards against errors/decisions~\cite{lubars2019ask, hall2017accountability, lerner1999accounting}. 
This could mean, rather than avoiding AI, developers strategically use it in high-accountability tasks.

\textbf{H3}. \textit{Higher task accountability increases developers’ openness to AI support and usage.}
However, accountability lowers automation bias~\cite{parasuraman2000model, parasuraman2010complacency}. Since mistakes ultimately fall on them, developers are likely to be cautious, insisting on oversight and decision control.

\textbf{Demands} capture the cognitive effort a task imposes~\cite{bakker2007job}. High-demand work strains coping resources, increasing receptivity to aids that reduce mental load~\cite{sweller1988cognitive, lazarus1991emotion}. Developers may use AI to lower cognitive costs and sustain momentum in demanding work.

\textbf{H4}. \textit{Higher task demands increase developers’ openness to AI support and usage.}

\textbf{Controls \& Groups:} We control for developers’ SE and AI experience, as both can shape baseline attitudes toward AI~\cite{crowston2025deskilling, bansal2021does}. 
Individual traits can also condition how appraisals translate into AI use. We emphasize \textit{risk tolerance} and \textit{technophilia}~\cite{burnett2016gendermag}—traits linked to stronger AI-adoption dispositions~\cite{choudhuri2025guides}. For example, risk-tolerant developers may feel less deterred by accountability pressures, while technophiles may actively seek AI integration opportunities~\cite{choudhuri2025guides}. We expect these traits to moderate the hypothesized relationships.

\section{Method} 
\label{sec:method}
To address our RQs, we surveyed software developers at Microsoft. 
Microsoft employs over 60{,}000 developers worldwide, spanning diverse domains, team structures, processes, and stakeholder contexts. This scale, combined with exposure to both mature and emerging AI tooling, makes it a rich and diverse setting for our study.

\subsection{Study Design} 

The goal of our study was to: (1) characterize how developers appraise SE tasks; (2) assess how these appraisals shape AI openness and use; (3) identify opportunities/gaps where AI can better support developer workflows; and (4) understand which Responsible AI (RAI) design principles developers prioritize in AI tools to credibly support SE tasks. The study was approved by Microsoft’s IRB.

\textbf{Synthesizing a taxonomy of SE tasks}: To study task appraisals (RQ1), we first constructed a representative, grounded taxonomy of SE tasks (Table~\ref{tab:categories}), integrating multiple empirical sources~\cite{kumar2025time, meyer2019today, choudhuri2025guides, khemka2024toward}. We drew on work-week studies of developer activities~\cite{kumar2025time, meyer2019today} for detailed task inventories and groupings. We then enriched this set with job-distribution insights from large-scale developer surveys on AI adoption~\cite{choudhuri2025guides, khemka2024toward}, ensuring our taxonomy reflected SE responsibilities across roles, geographies, and contexts. Finally, we triangulated coverage through pilot sessions with developers and SE researchers outside our team, identifying any missing tasks and validating the clarity of category boundaries.

\begin{table}[bhtp]
\small
\caption{Grounded taxonomy of SE tasks~\cite{kumar2025time, meyer2019today, choudhuri2025guides, khemka2024toward}}
\label{tab:categories}
\centering
\begin{tabular}{>{\raggedright\arraybackslash}m{2.3cm} >{\raggedright\arraybackslash}m{5.7cm}}
\hline
\textbf{Category} & \textbf{Tasks} \\
\hline \hline
\rowcolor{gray!25} Development & Coding/Programming, Bug Fixing/Debugging, Performance Optimization, Refactoring \& Maintenance/Updates, AI Integration\\ 
Design \& Planning & System Design, Requirements Engineering, Project Planning \& Management \\
\rowcolor{gray!25} Quality \& Risk Management & Testing \& Quality Assurance; Code Review/Pull Requests; Security \& Compliance \\
Infrastructure \& Operations & DevOps(CI/CD); Environment Setup \& Maintenance; Infrastructure Monitoring; Customer Support \\
\rowcolor{gray!25} Meta-work (Collaboration/Knowledge work) & Documentation; Client/Stakeholder Communication; Mentoring \& Onboarding; Learning; Research \& Brainstorming \\
\hline
\end{tabular}
\end{table}

\textbf{Responsible AI (RAI) principles}:
To assess developers’ RAI priorities (RQ2), we anchored our study in Microsoft’s Responsible AI framework~\cite{MicrosoftRAI2025}. This framework synthesizes established AI ethics guidelines~\cite{MontrealDeclaration2017, IBMTrustworthyAI2021, european2019ethics, GoogleAIPrinciples2023} and includes: \textit{Reliability \& Safety, Privacy \& Security, AI Accountability (provenance), Fairness, Inclusiveness,} and \textit{Transparency}. We extended this set with \textit{Steerability} (\textit{user agency/autonomy}) and \textit{Goal maintenance} (sustained \textit{alignment} with user goals), both emphasized in recent RAI research~\cite{Anthropic2025AgenticMisalignment, tankelevitch2024metacognitive, jakesch2022different, choudhuri2025guides}. 

\begin{table}[!ht]
\small
\caption{Theoretical constructs and instruments}
\label{tab:mm-instruments}
\centering
\begin{tabular}{>{\raggedright\arraybackslash}m{3cm} >{\raggedright\arraybackslash}m{5cm}}
\hline
\textbf{Construct} & \textbf{Instrument} \\
\hline \hline
\rowcolor{gray!25} Value & Job Characteristics Model~\cite{fried1987validity, trinkenreich2024predicting} \\ 
Identity & Self-Determination Theory~\cite{venkatesh2012consumer} \\ 
\rowcolor{gray!25} Accountability & Felt Accountability Scale~\cite{hall2017accountability} \\ 
Demands & Job Demands-Resource Model~\cite{bakker2007job}\\
\rowcolor{gray!25} Openness to AI Support & Levels of Automation Framework \cite{shao2025future, parasuraman2000model} \\ 
 AI Usage & Technology Acceptance (UTAUT) \cite{venkatesh2012consumer} \\ 
\rowcolor{gray!25} Risk Tolerance, Technophilia & Cognitive Style Facet Survey \cite{choudhuri2025guides} \\ 

\hline
\end{tabular}
\end{table}

\textbf{Survey design}: 
We followed Kitchenham’s guidelines for conducting surveys~\cite{kitchenham2008personal} and drew on established theoretical frameworks and validated instruments from behavioral sciences and Human–AI Interaction (HAI) research (Table~\ref{tab:mm-instruments}).
The survey was refined through iterative validation with external researchers and multiple pilot rounds. It comprised three sections:  

(1) \textit{AI experience and dispositions}: After informed consent, participants reported their experience with AI tools and their dispositions toward its use in work. We prefaced this section with a description of developer-facing AI tools from the DORA 2025 survey~\cite{dora2025}. Participants with no AI-tool experience exited at this point.

(2) \textit{Background \& Demographics}: Participants reported SE experience and, optionally, gender and country of residence. They then selected 2-3 task categories (in Table~\ref{tab:categories}) that best reflected their current work and answered the subsequent questions for those categories. To reduce fatigue, the meta-work category (applicable to all developers) was excluded from the initial selection and shown only if a participant had selected two categories; thus, no participant completed more than three category blocks.

(3) \textit{Task Category blocks}: Each task category was a separate block. For each selected task category (see Tab.~\ref{tab:categories}), participants answered:

\quad\textit{(a) RQ1: Task appraisals and AI use.} For each task in a category (e.g., Testing/QA, Security, Code Review under Quality \& Risk Management), participants rated task \textit{value, identity, accountability}, and \textit{demands}, captured using validated instruments (Table~\ref{tab:mm-instruments}). We used single-item measures to reduce participant fatigue, as these retain psychometric validity for well-scoped constructs~\cite{matthews2022normalizing}. Participants then reported their \textit{openness to AI support} and \textit{AI use} frequency for each task (dependent variables in RQ1). Finally, they answered two open-ended questions: \textit{where they most wanted AI support}, and \textit{where they preferred to limit it}; within the category and why.

\quad\textit{(b) RQ2—RAI priorities.} Participants selected any five RAI principles (from the eight listed earlier) they deemed most important for AI tooling in that category (the five-choice format was drawn from~\cite{jakesch2022different}). This forces trade-offs and mitigates ceiling effects (``all-high'' bias common in Likert importance ratings)~\cite{alwin1991reliability, bradburn2004asking}. After selecting, participants could optionally describe experiences that informed their choices. We tested alternative elicitation formats (ranking, sorting, point allocation, MaxDiff)~\cite{bradburn2004asking} and chose this approach based on sandbox feedback. Since RAI principles can be abstract and difficult to contextually realize~\cite{cave2018portrayals}, we provided on-demand plain-language explanations (adapted from~\cite{jakesch2022different}) via information icons next to each principle. Each explanation included: (a) what a system embodying the principle would do, and (b) an example realizing its application, while retaining a degree of generality (see~\cite{jakesch2022different}). 

In the pilot, participants could suggest missing tasks (for categories they answered) and/or provide other feedback.

%

We administered the survey in Qualtrics \cite{qualtrics2025}. All closed-ended questions used a 5-point Likert scale, with a sixth option (“I’m not sure” or “I don’t do this task/N.A.”) to distinguish ignorance from indifference \cite{grichting1994meaning}. The survey took 10–15 minutes to complete. To ensure data quality and reduce response bias, we included attention checks, randomized questions within blocks, and the order of task-category blocks. The survey instrument is in supplemental~\cite{supplemental}.

\textbf{Sandbox and pilot}: We sandboxed the survey with developers and SE/HCI researchers (n=11) to assess clarity, interpretability, and realism. Based on their feedback, we added safeguards against automated submissions and revised questions to reflect participants’ current work.
%
Initially, all participants saw the meta-work block (Table~\ref{tab:categories}), but pilots showed that limiting respondents to three category blocks improved data quality and reduced fatigue. We also tested multiple elicitation formats for RAI prioritization~\cite{bradburn2004asking}. Respondents preferred selecting top-N principles and explaining tradeoffs over ranking ethical values. We adopted this format consistent with prior work~\cite{jakesch2022different}. 
To finalize the survey, we piloted it with (n=50) developers, validating clarity and task coverage. Minor wording edits were made; pilot responses were excluded from analysis.


\subsection{\textbf{Data Collection}} 
\textbf{Distribution}: We distributed the survey to 8{,}000 software developers at Microsoft via email in July 2025, sampled uniformly at random across product groups, roles, and geographies. To incentivize participation, respondents could enter a raffle for ten~\$50 AmEx gift cards. One reminder was sent after a week. Participation was voluntary; responses were anonymous unless participants opted into follow-up contact.

\textbf{Sample size}: To determine the appropriate sample size, we conducted an a priori power analysis in G*Power~\cite{faul2009statistical} for multiple linear regression with repeated measures, using the number of predictors in our design. We targeted the detection of even a small effect size ($d = 0.05$) at a significance level of $\alpha = 0.05$ with power = 0.95. The analysis indicated a minimum of 245 responses. To accommodate missing data, quality exclusions, and subgroup analyses, we targeted at least three times this number.

\textbf{Responses}: We received 1{,}193 responses, a response rate of 14.86\% (consistent with prior SE surveys~\cite{storey2019towards, punter2003conducting}). We removed incomplete ($n = 152$) and patterned responses (straight-lined or repetitive altering; $n = 59$), as well as those that failed attention checks ($n = 98$) or reported no AI experience ($n = 24$). We treated \textit{``I’m not sure’’/``I don’t do this task''} Likert selections as missing data. 

We retained 860 valid responses from developers across 6 continents, representing wide distributions of SE and AI experience. Most respondents were from North America (57.4\%) and identified as men (73.8\%), consistent with distributions reported in prior SE studies~\cite{trinkenreich2023model, choudhuri2025guides, russo2024navigating}. 
Demographics are summarized in~\cite{supplemental}.

\subsection{Data Analysis}
\label{instr-design}

\textbf{Quantitative}: We analyzed data in Python and R to summarize distributions, fit regression models (see \textsection~\ref{rq1-hyp},\ref{rq2}), and generate visualizations. Closed-ended responses (Likert; Top-N) were visualized to assess variation in appraisals, AI openness and usage across tasks (Tab.~\ref{tab:task_drivers}, Fig.~\ref{fig:rq1-plots}) and RAI priorities across categories (Tab.~\ref{tab:regression-haiaspects}, Fig.~\ref{fig:design_principle}).
For RQ1, the unit of analysis was \emph{(participant, task type)}; for RQ2, \emph{(participant, task category)}. Because the design involved repeated measures within participants \& across tasks, we used mixed-effects regression~\cite{gelman2007data}. Model specifications appear in \S\ref{sec:Res}.

\textbf{Qualitative}: We analyzed free-text responses using reflexive thematic analysis~\cite{braun2006using, braun2022conceptual}. To ensure rigor, the team held multiple meetings to iteratively compare interpretations, refine codes, and build shared understanding, following established practices~\cite{creswell2016qualitative, braun2022conceptual}.

First, we inductively open-coded the data, then refined and consolidated codes, merging conceptually similar ones while keeping others distinct. Team discussions relied on negotiated agreement to reach consensus on final themes (cataloged in~\cite{supplemental}). 
Next, to understand why patterns emerged, we mapped qualitative themes to quantitative results through joint interpretation, comparing participants’ free-text responses with their Likert-scale selections. We observed no inconsistencies between their assessments and explanations.  Where relevant, we triangulated findings with theory to structure interpretation.

We analyzed 1,528 responses about where developers seek and limit AI support and 2,453 responses explaining RAI priorities, spanning five task categories. Participants are referenced as P1–P860.

Finally, we validated our findings through member checking: findings were shared with 371 participants who opted into follow-up, and 62 responded. Their feedback affirmed the interpretations and offered clarifications; no new insights or disagreements emerged.

\section{Results}
\label{sec:Res}

This section reports (1) how developers' task appraisals shape AI adoption (\ref{rq1-hyp}), (2) where they seek or limit AI support (\ref{want-AI}), and (3) the RAI principles they prioritize in AI support for SE tasks (\ref{rq2}).

\subsection{RQ1a: How do appraisals shape AI adoption?}
\label{rq1-hyp}
To answer RQ1, we first investigated whether developers' task appraisals predicted their (a) openness to AI support and (b) AI usage (outcomes), and whether these relationships varied by developers' experience and AI dispositions.

For each outcome, we fit linear mixed-effects regressions~\cite{gelman2007data}, with appraisals as fixed effects; controls for developers’ SE and AI experience, and random effects for within-\textit{participant} and across-\textit{task type} dependence. 
Models were estimated for the full sample (Tab.~\ref{tab:hyp-coeff}) and, per our planned group analyses, stratified by \textit{risk tolerance} and \textit{technophilic motivations} (recall ~\textsection~\ref{sec:theory}). Group analyses statistics are in~\cite{supplemental}. All Variance Inflation Factors (VIFs) were~$<2$ \cite{hair2009multivariate}, confirming lack of multicollinearity concerns. We controlled false discovery rates (FDR) using Benjamini–Hochberg~\cite{thissen2002quick} and report statistical significance after this correction.


\begin{table}[ht]
\centering
\small
\caption{\small Mixed-effects regression results for developers' (a) openness to AI support and (b) AI usage, estimated for the full sample (N = 860). Cells report standardized regression coefficients ($\beta$), p-values, and effect sizes (d) in parentheses. Blank cells indicate non-significant associations after Benjamini-Hochberg FDR adjustment~\cite{thissen2002quick}.}
\begin{tabular}{lcc}
\toprule
\toprule
\textbf{Factor} & \textbf{Openness to AI support} & \textbf{Reported AI usage} \\
\midrule

Value (H1)  & .12*** (.16)  & .16*** (.18) \\
Identity (H2) & -.09*** (-.15)  & .15*** (.20) \\
Accountability (H3) & .07*** (.10) & .18*** (.21) \\
Demand (H4) & .12*** (.18)  & .09*** (.10) \\
  \midrule
SE Experience &    & -.09*** (-.13) \\
AI Experience & .19*** (.27) & .41*** (.46) \\
\midrule
\ $R^2_m$ /\ $R^2_c$ & .25 /.45  & .25 /.48 \\
Observations & \multicolumn{2}{c}{10,449} \\
\bottomrule
\end{tabular}
\begin{tablenotes}
\item \small {\textit{$^{*}p<.05$; $^{**}p<.01$; $^{***}p<.001$.} \textit{We consider $d<$ 0.02 to be no effect, $d\in$ [0.02, 0.15) small, $d \in$ [0.15, 0.35) medium, and $d >$ 0.35 large \cite{cohen2013statistical}}}
\end{tablenotes}
\label{tab:hyp-coeff}
\end{table}

Table~\ref{tab:hyp-coeff} summarizes regression results. All hypothesized effects (H1–H4) were supported. We report marginal ($R^2_m$; variance from fixed effects) and conditional ($R^2_c$; variance from fixed and random effects) fit indices as indicators of model fit~\cite{hair2009multivariate}.


\textbf{Task Value (H1)} positively predicted openness to and use of AI support. A one-standard deviation (SD) increase in perceived value raised openness by .12 and use by .16 SD units (p < 0.001, FDR-corrected), with medium effects (.16, .18) holding other factors constant (Table~\ref{tab:hyp-coeff}). When developers viewed tasks as important, they used AI to boost efficiency (e.g., seeking assistance in tedious steps). They, nonetheless, stressed retaining decision control~(see~\textsection\ref{want-AI}), positioning AI as complementary rather than substitutive. 

\textbf{Task Identity (H2)} alignment showed a dual pattern: lower openness to AI support ($\beta$=-.09, p<.001) but higher usage ($\beta$=.15, p<.001), with medium effects (-.15, .20). Developers protected ownership of identity-defining work; yet used AI to refine their craft (e.g., learning, research, exploration) (see ~\textsection\ref{want-AI}).

\textbf{Task Accountability (H3)} positively associated with openness ($\beta=.07$, $p<.001$) and use ($\beta=.18$, $p<.001$), with small–medium effects (.10, .21). Developers leveraged AI as a safeguard in high-stakes tasks. Yet heightened accountability increased vigilance: they insisted on deliberate review, maintained oversight, and retained decision control for these tasks (see~\textsection\ref{want-AI}).


\textbf{Task Demands (H4)} positively associated with openness ($\beta$=.12, p<.001) and use ($\beta$=.09, p<.001), with small-medium effects (.18, .10). For demanding work, developers used AI to offload rote steps, lower cognitive load, and sustain momentum (see~\textsection\ref{want-AI}). In such cases, AI functioned as a cognitive scaffold that freed attention for higher-order knowledge work~\cite{lee2025impact}.

\textbf{Experience:} SE experience predicted lower AI use ($\beta=-.09$, p<.001). Experienced developers are likely to rely on established repertoires~\cite{ericsson2006influence}, reducing the perceived utility of AI delegation; while juniors use AI to offset skill gaps~\cite{crowston2025deskilling}. Openness did not differ significantly by SE experience. Meanwhile, prior AI experience increased both AI openness and use ($\beta=.19,.41$; p<.001).


\textbf{Group analysis (AI Dispositions):} Stratifying by median splits on AI dispositions, risk-tolerant developers showed higher openness and use overall. They sought more AI support for high-value ($\Delta\beta=.06$, $p=.035$) and high-demand ($\Delta\beta=.09$, $p=.001$) tasks, and used AI more in high-accountability ($\Delta\beta=.07$, $p=.038$) and demanding scenarios ($\Delta\beta=.08$, $p=.002$). Risk-averse peers remained more vigilant under accountability pressures~\cite{tetlock1983accountability,lerner1999accounting}. 

Technophiles also showed higher openness and use overall. Critically, accountability appraisals (H3) predicted these outcomes only among high-technophiles (Openness: $\beta=0.07$; Usage: $\beta=0.20$; both $p<.001$) but not among low-technophiles, indicating that technophily moderates AI adoption in high-stakes conditions. High-technophiles have the orchestration habits to use AI as a “second set of eyes”~\cite{bansal2021does}, while low-technophiles---lacking these routines---view AI as net-costly under accountability pressure. Other associations were consistent across groups (see supplemental~\cite{supplemental}).

\vspace{2mm}


\begin{takeawayBox}

\textbf{Takeaway}: Appraisals shape AI adoption. Task \textit{Value}, \textit{Accountability}, and \textit{Demands} increase AI openness and use, while \textit{Identity}-alignment shows dual effects (lower openness, higher use). 
Junior, AI-experienced, risk-tolerant, and technophilic developers show greater receptivity overall.

\end{takeawayBox}

\subsection{RQ1b: Where and why do developers seek or limit AI support?}
\label{want-AI}

We examined how appraisals varied across tasks to locate \textit{where} and \textit{why} developers seek or limit AI support. 

To do so, we first clustered tasks by their appraisal signatures (Tab.~\ref{tab:task_drivers}). For each task, we computed top-2 agreement proportions (share selecting 4–5 on a 5-point scale~\cite{ladhari2010developing}) for the four appraisals, then standardized these values to z-scores ($z=(x-\bar{x})/\mathrm{sd}(x)$) for cross-scale comparability~\cite{hair2009multivariate}. We applied agglomerative hierarchical clustering (Ward linkage)~\cite{ward1963hierarchical} on Euclidean distances of these z-scores, selecting $k=3$ optimal clusters via silhouette analysis~\cite{rousseeuw1987silhouettes} (see~\cite{supplemental} for silhouette plot).
We used precision-weighting (inverse-variance shrinkage) to address unequal task-Ns and validated cluster stability via stratified bootstraps (B=1000)~\cite{hair2009multivariate}. 

The analysis yielded the following clusters:


\begin{itemize}[leftmargin=1em]
\item \textbf{C1: Core work} — High value and demands; moderate–high accountability; moderate–strong identity alignment. 
\item \textbf{C2: People \& AI-building} — Moderate value, demands, and accountability; strong identity alignment. 
\item \textbf{C3: Ops \& Coordination} — Moderate–high value, demands, and accountability; weak identity alignment. 
\end{itemize}

\begin{figure}[!thb]
\small
\centering
\includegraphics[width=0.48
\textwidth]{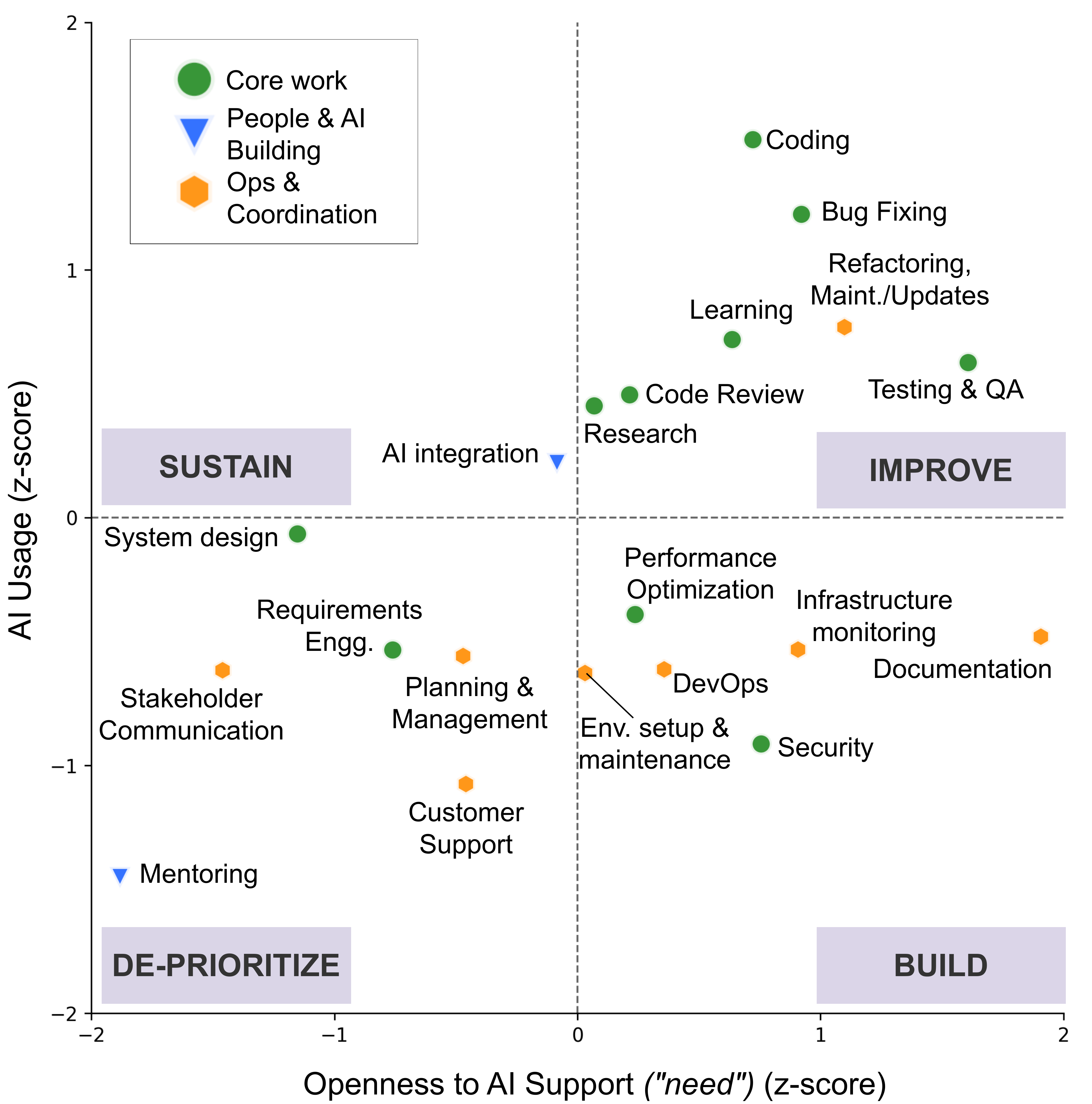}
\caption{Scatter z-score plot showing relative AI-support vs Usage ($z$) scores for SE tasks. Tasks are grouped into four quadrants: Build, Improve, Sustain, and De-prioritize.}
\label{fig:rq1-plots}
\end{figure}

We simultaneously mapped tasks onto an \textit{Openness to AI Support (x) vs. AI Usage (y)} plane (Fig.~\ref{fig:rq1-plots}) to visualize gaps in tooling support. Axes show task-level z-scores for openness ``need'' (x) and reported use (y). Quadrants (mean-split: z=0) highlight distinct gaps:

\begin{itemize}[leftmargin=1em]
  \item \textbf{Build} (bottom-right; high need, low use): Clear need but limited adoption; reduce friction and prototype new support. 

  \item \textbf{Improve} (top-right; high need, high use): Strong need and adoption; focus on reliability and quality for gains.
  
  \item \textbf{Sustain} (top-left; low need, high use): AI is used but not essential; maintain support without over-investment.
  
  \item \textbf{De-prioritize} (bottom-left; low need, low use): Limited uptake; expect lower returns from additional investment.
\end{itemize}

\newlength{\myDistLength}
\setlength{\myDistLength}{25pt}

\def\myfactor#1{#1}

\def\mybarchartcounts#1#2#3#4#5{%
  \pgfmathsetmacro{\totalcount}{#1 + #2 + #3 + #4 + #5}%
  \pgfmathsetmacro{\propone}{#1 / \totalcount}%
  \pgfmathsetmacro{\proptwo}{#2 / \totalcount}%
  \pgfmathsetmacro{\propthree}{#3 / \totalcount}%
  \pgfmathsetmacro{\propfour}{#4 / \totalcount}%
  \pgfmathsetmacro{\propfive}{#5 / \totalcount}%
  \resizebox{\the\myDistLength}{7.5pt}{%
    \begin{tikzpicture}
      \begin{axis}[
        axis background/.style={fill=gray!10, draw=gray!50},
        axis line style={draw=none},
        tick style={draw=none},
        ytick=\empty,
        xtick=\empty,
        ymin=0, ymax=1.0,
        xmin=0, xmax=5]
        \addplot[
          ybar interval=.5,
          fill=TekheletPurple,
          draw=none,
        ] coordinates {(5,\propfive) (4,\propfour) (3,\propthree) (2,\proptwo) (1,\propone)};
      \end{axis}
    \end{tikzpicture}%
  }%
}

\definecolor{Gradient314}{HTML}{F8ED81}
\definecolor{Gradient350}{HTML}{F8ED81}
\definecolor{Gradient406}{HTML}{F5EB81}
\definecolor{Gradient419}{HTML}{EEE981}
\definecolor{Gradient483}{HTML}{CEDE83}
\definecolor{Gradient508}{HTML}{C1DA84}
\definecolor{Gradient516}{HTML}{BDD884}
\definecolor{Gradient519}{HTML}{BBD884}
\definecolor{Gradient574}{HTML}{9ECD86}
\definecolor{Gradient590}{HTML}{95CA86}
\definecolor{Gradient599}{HTML}{90C986}
\definecolor{Gradient619}{HTML}{83C490}
\definecolor{Gradient622}{HTML}{81C391}
\definecolor{Gradient648}{HTML}{70BD9E}
\definecolor{Gradient665}{HTML}{65BAA6}
\definecolor{Gradient671}{HTML}{61B8A9}
\definecolor{Gradient673}{HTML}{5FB8AA}
\definecolor{Gradient674}{HTML}{5FB7AB}
\definecolor{Gradient675}{HTML}{5EB7AB}
\definecolor{Gradient678}{HTML}{5CB7AD}
\definecolor{Gradient688}{HTML}{55B4B2}
\definecolor{Gradient700}{HTML}{4EB2B8}
\definecolor{Gradient715}{HTML}{4BABB8}
\definecolor{Gradient724}{HTML}{49A7B8}
\definecolor{Gradient726}{HTML}{49A6B8}
\definecolor{Gradient734}{HTML}{48A3B8}
\definecolor{Gradient740}{HTML}{47A0B8}
\definecolor{Gradient742}{HTML}{469FB8}
\definecolor{Gradient746}{HTML}{469DB8}
\definecolor{Gradient747}{HTML}{469DB8}
\definecolor{Gradient749}{HTML}{459CB8}
\definecolor{Gradient750}{HTML}{459CB8}
\definecolor{Gradient755}{HTML}{4499B8}
\definecolor{Gradient759}{HTML}{4398B8}
\definecolor{Gradient763}{HTML}{4396B8}
\definecolor{Gradient766}{HTML}{4294B8}
\definecolor{Gradient771}{HTML}{4192B8}
\definecolor{Gradient793}{HTML}{3E89B8}
\definecolor{Gradient804}{HTML}{3B82B5}
\definecolor{Gradient805}{HTML}{3B81B4}
\definecolor{Gradient807}{HTML}{3A80B3}
\definecolor{Gradient810}{HTML}{397DB1}
\definecolor{Gradient823}{HTML}{3572A8}
\definecolor{Gradient825}{HTML}{3470A7}
\definecolor{Gradient838}{HTML}{30659E}
\definecolor{Gradient839}{HTML}{30649D}
\definecolor{Gradient843}{HTML}{2E619B}
\definecolor{Gradient844}{HTML}{2E609A}
\definecolor{Gradient854}{HTML}{2B5893}
\definecolor{Gradient857}{HTML}{2A5591}
\definecolor{Gradient861}{HTML}{28528F}
\definecolor{Gradient864}{HTML}{274F8D}
\definecolor{Gradient868}{HTML}{264C8A}
\definecolor{Gradient876}{HTML}{234585}
\definecolor{Gradient878}{HTML}{234383}
\definecolor{Gradient888}{HTML}{1F3B7D}
\definecolor{Gradient889}{HTML}{1F3A7C}
\definecolor{Gradient895}{HTML}{1D3578}
\definecolor{Gradient903}{HTML}{1C3175}
\definecolor{Gradient907}{HTML}{1C3175}
\definecolor{Gradient909}{HTML}{1C3175}
\definecolor{Gradient911}{HTML}{1C3175}
\definecolor{Gradient913}{HTML}{1C3175}
\definecolor{Gradient923}{HTML}{1C3175}
\definecolor{Gradient934}{HTML}{1C3175}
\definecolor{Gradient935}{HTML}{1C3175}
\definecolor{Gradient949}{HTML}{1C3175}
\definecolor{Gradient955}{HTML}{1C3175}
\definecolor{Gradient958}{HTML}{1C3175}
\definecolor{Gradient967}{HTML}{1C3175}
\definecolor{Gradient968}{HTML}{1C3175}
\definecolor{Gradient969}{HTML}{1C3175}
\definecolor{Gradient970}{HTML}{1C3175}
\definecolor{Gradient979}{HTML}{1C3175}
\definecolor{Gradient980}{HTML}{1C3175}

\begin{table*}[ht]
\centering
  \caption{\small Task appraisal profiles across four drivers: Value, Identity, Accountability, and Demand. Tasks are grouped into three clusters derived from driver-response patterns: \emph{Core Work}, \emph{People \& AI Building}, and \emph{Ops \& Coordination}. Within each cluster, tasks are sorted by Value agreement (\% Agree/Strongly Agree). For each driver, we show (i) the full 5-point Likert distribution, (ii) the \% Agree/Strongly Agree with a common color gradient for comparison, and (iii) the task’s rank on that driver. The legend spans 0\% (yellow) to 100\% (dark blue). Confidence intervals for percentages are in the supplemental; tasks with overlapping CIs are not statistically distinguishable in rank.}
{\small
\begin{tabular}{@{} l ccc ccc ccc ccc @{\hspace{4mm}} p{30pt} @{}}
\toprule
\toprule
 & \multicolumn{3}{c}{\textbf{Task Value}} & \multicolumn{3}{c}{\textbf{Task Identity}} & \multicolumn{3}{c}{\textbf{Task Accountability}} & \multicolumn{3}{c}{\textbf{Task Demand}}\\
\cmidrule(lr){2-4}\cmidrule(lr){5-7}\cmidrule(lr){8-10}\cmidrule(lr){11-13}
\textbf{Task} & Dist. & \% & Rnk & Dist. & \% & Rnk & Dist. & \% & Rnk & Dist. & \% & Rnk\\
& \footnotesize\makebox[\the\myDistLength][s]{\color{TekheletPurple}1\hfill 2\hfill 3\hfill 4\hfill 5} & & & \footnotesize\makebox[\the\myDistLength][s]{\color{TekheletPurple}1\hfill 2\hfill 3\hfill 4\hfill 5} & & & \footnotesize\makebox[\the\myDistLength][s]{\color{TekheletPurple}1\hfill 2\hfill 3\hfill 4\hfill 5} & & & \footnotesize\makebox[\the\myDistLength][s]{\color{TekheletPurple}1\hfill 2\hfill 3\hfill 4\hfill 5} & & \\
\midrule
\multicolumn{13}{l}{\textbf{Cluster 1 --- Core Work}}\\
\myfactor{Coding/Programming} & \mybarchartcounts{1}{3}{12}{138}{634} & \cellcolor{Gradient980}{\color{white}98.0} & \#1 & \mybarchartcounts{0}{2}{22}{211}{554} & \cellcolor{Gradient970}{\color{white}97.0} & \#1 & \mybarchartcounts{0}{10}{30}{177}{572} & \cellcolor{Gradient949}{\color{white}94.9} & \#1 & \mybarchartcounts{6}{45}{130}{343}{265} & \cellcolor{Gradient771}{\color{white}77.1} & \#13 & \tikzmark{legend_top} \\
\myfactor{System Design} & \mybarchartcounts{0}{3}{8}{135}{378} & \cellcolor{Gradient979}{\color{white}97.9} & \#2 & \mybarchartcounts{1}{6}{39}{162}{320} & \cellcolor{Gradient913}{\color{white}91.3} & \#3 & \mybarchartcounts{1}{14}{50}{170}{289} & \cellcolor{Gradient876}{\color{white}87.6} & \#3 & \mybarchartcounts{1}{7}{41}{221}{258} & \cellcolor{Gradient907}{\color{white}90.7} & \#1 &  \\
\myfactor{Testing \& QA} & \mybarchartcounts{1}{1}{10}{97}{275} & \cellcolor{Gradient969}{\color{white}96.9} & \#3 & \mybarchartcounts{9}{48}{100}{151}{75} & \cellcolor{Gradient590}{\color{black}59.0} & \#11 & \mybarchartcounts{0}{15}{44}{134}{186} & \cellcolor{Gradient844}{\color{white}84.4} & \#5 & \mybarchartcounts{1}{6}{43}{161}{168} & \cellcolor{Gradient868}{\color{white}86.8} & \#5 &  \\
\myfactor{Bug Fixing/Debugging} & \mybarchartcounts{2}{3}{20}{184}{581} & \cellcolor{Gradient968}{\color{white}96.8} & \#4 & \mybarchartcounts{16}{68}{126}{298}{281} & \cellcolor{Gradient734}{\color{white}73.4} & \#8 & \mybarchartcounts{4}{14}{43}{196}{532} & \cellcolor{Gradient923}{\color{white}92.3} & \#2 & \mybarchartcounts{3}{26}{84}{329}{346} & \cellcolor{Gradient857}{\color{white}85.7} & \#6 &  \\
\myfactor{Code Review/Pull Requests} & \mybarchartcounts{1}{2}{10}{109}{267} & \cellcolor{Gradient967}{\color{white}96.7} & \#5 & \mybarchartcounts{4}{23}{64}{159}{139} & \cellcolor{Gradient766}{\color{white}76.6} & \#5 & \mybarchartcounts{0}{13}{41}{124}{210} & \cellcolor{Gradient861}{\color{white}86.1} & \#4 & \mybarchartcounts{2}{22}{77}{180}{108} & \cellcolor{Gradient740}{\color{white}74.0} & \#16 &  \\
\myfactor{Requirements Engineering} & \mybarchartcounts{0}{3}{19}{189}{308} & \cellcolor{Gradient958}{\color{white}95.8} & \#6 & \mybarchartcounts{13}{48}{112}{227}{116} & \cellcolor{Gradient665}{\color{white}66.5} & \#10 & \mybarchartcounts{6}{29}{95}{187}{197} & \cellcolor{Gradient747}{\color{white}74.7} & \#11 & \mybarchartcounts{1}{8}{49}{255}{205} & \cellcolor{Gradient888}{\color{white}88.8} & \#3 &  \\
\myfactor{Security \& Compliance} & \mybarchartcounts{1}{1}{15}{91}{267} & \cellcolor{Gradient955}{\color{white}95.5} & \#7 & \mybarchartcounts{27}{62}{89}{117}{73} & \cellcolor{Gradient516}{\color{black}51.6} & \#14 & \mybarchartcounts{0}{22}{42}{102}{195} & \cellcolor{Gradient823}{\color{white}82.3} & \#6 & \mybarchartcounts{0}{16}{48}{138}{163} & \cellcolor{Gradient825}{\color{white}82.5} & \#8 &  \\
\myfactor{Research \& Brainstorming} & \mybarchartcounts{0}{7}{36}{182}{245} & \cellcolor{Gradient909}{\color{white}90.9} & \#10 & \mybarchartcounts{0}{11}{41}{199}{217} & \cellcolor{Gradient889}{\color{white}88.9} & \#4 & \mybarchartcounts{2}{24}{89}{172}{183} & \cellcolor{Gradient755}{\color{white}75.5} & \#9 & \mybarchartcounts{0}{14}{48}{196}{212} & \cellcolor{Gradient868}{\color{white}86.8} & \#4 &  \\
\myfactor{Performance Optimization} & \mybarchartcounts{1}{16}{76}{272}{399} & \cellcolor{Gradient878}{\color{white}87.8} & \#12 & \mybarchartcounts{15}{55}{121}{282}{292} & \cellcolor{Gradient750}{\color{white}75.0} & \#6 & \mybarchartcounts{8}{43}{130}{212}{358} & \cellcolor{Gradient759}{\color{white}75.9} & \#8 & \mybarchartcounts{2}{19}{59}{273}{410} & \cellcolor{Gradient895}{\color{white}89.5} & \#2 &  \\
\myfactor{Learning} & \mybarchartcounts{3}{12}{62}{186}{215} & \cellcolor{Gradient839}{\color{white}83.9} & \#16 & \mybarchartcounts{0}{6}{25}{189}{257} & \cellcolor{Gradient935}{\color{white}93.5} & \#2 & \mybarchartcounts{7}{34}{90}{150}{197} & \cellcolor{Gradient726}{\color{white}72.6} & \#13 & \mybarchartcounts{2}{23}{66}{207}{180} & \cellcolor{Gradient810}{\color{white}81.0} & \#9 &  \\
\midrule
\multicolumn{13}{l}{\textbf{Cluster 2 --- People \& AI Building}}\\
\myfactor{Mentoring \& Onboarding} & \mybarchartcounts{3}{29}{105}{173}{129} & \cellcolor{Gradient688}{\color{white}68.8} & \#19 & \mybarchartcounts{7}{28}{96}{186}{120} & \cellcolor{Gradient700}{\color{white}70.0} & \#9 & \mybarchartcounts{12}{38}{114}{136}{134} & \cellcolor{Gradient622}{\color{black}62.2} & \#18 & \mybarchartcounts{4}{44}{95}{197}{94} & \cellcolor{Gradient671}{\color{white}67.1} & \#20 &  \\
\myfactor{AI Integration} & \mybarchartcounts{25}{52}{133}{196}{238} & \cellcolor{Gradient674}{\color{white}67.4} & \#20 & \mybarchartcounts{20}{33}{103}{236}{231} & \cellcolor{Gradient750}{\color{white}75.0} & \#7 & \mybarchartcounts{41}{60}{141}{161}{200} & \cellcolor{Gradient599}{\color{black}59.9} & \#20 & \mybarchartcounts{6}{29}{140}{232}{207} & \cellcolor{Gradient715}{\color{white}71.5} & \#18 &  \\
\midrule
\multicolumn{13}{l}{\textbf{Cluster 3 --- Ops \& Coordination}}\\
\myfactor{DevOps (CI/CD)} & \mybarchartcounts{0}{3}{14}{92}{149} & \cellcolor{Gradient934}{\color{white}93.4} & \#8 & \mybarchartcounts{20}{45}{64}{80}{53} & \cellcolor{Gradient508}{\color{black}50.8} & \#15 & \mybarchartcounts{1}{14}{50}{88}{103} & \cellcolor{Gradient746}{\color{white}74.6} & \#12 & \mybarchartcounts{2}{23}{41}{112}{78} & \cellcolor{Gradient742}{\color{white}74.2} & \#15 &  \\
\myfactor{Infrastructure Monitoring} & \mybarchartcounts{1}{7}{15}{79}{157} & \cellcolor{Gradient911}{\color{white}91.1} & \#9 & \mybarchartcounts{29}{47}{58}{81}{44} & \cellcolor{Gradient483}{\color{black}48.3} & \#16 & \mybarchartcounts{4}{12}{47}{81}{107} & \cellcolor{Gradient749}{\color{white}74.9} & \#10 & \mybarchartcounts{0}{13}{37}{113}{92} & \cellcolor{Gradient804}{\color{white}80.4} & \#10 &  \\
\myfactor{Planning \& Management} & \mybarchartcounts{0}{7}{42}{212}{246} & \cellcolor{Gradient903}{\color{white}90.3} & \#11 & \mybarchartcounts{32}{83}{126}{168}{92} & \cellcolor{Gradient519}{\color{black}51.9} & \#13 & \mybarchartcounts{7}{41}{114}{165}{168} & \cellcolor{Gradient673}{\color{white}67.3} & \#16 & \mybarchartcounts{3}{21}{80}{252}{146} & \cellcolor{Gradient793}{\color{white}79.3} & \#11 &  \\
\myfactor{Refactoring \& Maintenance} & \mybarchartcounts{4}{12}{90}{298}{375} & \cellcolor{Gradient864}{\color{white}86.4} & \#13 & \mybarchartcounts{23}{120}{188}{255}{191} & \cellcolor{Gradient574}{\color{black}57.4} & \#12 & \mybarchartcounts{3}{28}{118}{224}{399} & \cellcolor{Gradient807}{\color{white}80.7} & \#7 & \mybarchartcounts{6}{58}{121}{308}{289} & \cellcolor{Gradient763}{\color{white}76.3} & \#14 &  \\
\myfactor{Env. Setup \& Maintenance} & \mybarchartcounts{0}{8}{29}{89}{133} & \cellcolor{Gradient857}{\color{white}85.7} & \#14 & \mybarchartcounts{22}{63}{66}{78}{31} & \cellcolor{Gradient419}{\color{black}41.9} & \#17 & \mybarchartcounts{4}{15}{70}{71}{93} & \cellcolor{Gradient648}{\color{white}64.8} & \#17 & \mybarchartcounts{0}{20}{50}{100}{84} & \cellcolor{Gradient724}{\color{white}72.4} & \#17 &  \\
\myfactor{Documentation} & \mybarchartcounts{1}{14}{54}{228}{176} & \cellcolor{Gradient854}{\color{white}85.4} & \#15 & \mybarchartcounts{44}{147}{133}{103}{45} & \cellcolor{Gradient314}{\color{black}31.4} & \#20 & \mybarchartcounts{6}{33}{108}{173}{151} & \cellcolor{Gradient688}{\color{white}68.8} & \#14 & \mybarchartcounts{0}{34}{64}{192}{183} & \cellcolor{Gradient793}{\color{white}79.3} & \#12 &  \\
\myfactor{Customer Support} & \mybarchartcounts{2}{7}{23}{62}{104} & \cellcolor{Gradient838}{\color{white}83.8} & \#17 & \mybarchartcounts{27}{36}{57}{55}{27} & \cellcolor{Gradient406}{\color{black}40.6} & \#18 & \mybarchartcounts{6}{22}{36}{59}{74} & \cellcolor{Gradient675}{\color{white}67.5} & \#15 & \mybarchartcounts{2}{6}{23}{81}{85} & \cellcolor{Gradient843}{\color{white}84.3} & \#7 &  \\
\myfactor{Stakeholder Communication} & \mybarchartcounts{0}{21}{61}{160}{179} & \cellcolor{Gradient805}{\color{white}80.5} & \#18 & \mybarchartcounts{31}{100}{142}{107}{40} & \cellcolor{Gradient350}{\color{black}35.0} & \#19 & \mybarchartcounts{13}{43}{104}{123}{137} & \cellcolor{Gradient619}{\color{black}61.9} & \#19 & \mybarchartcounts{3}{29}{104}{162}{124} & \cellcolor{Gradient678}{\color{white}67.8} & \#19 & \tikzmark{legend_bottom} \\
\bottomrule
\end{tabular}

\begin{tikzpicture}[overlay, remember picture]
    \definecolor{GradientYellow}{RGB}{248,237,129}
    \definecolor{GradientLightGreen}{RGB}{198,220,132}
    \definecolor{GradientGreen}{RGB}{144,201,135}
    \definecolor{GradientTeal}{RGB}{78,178,184}
    \definecolor{GradientCyan}{RGB}{61,134,184}
    \definecolor{GradientDarkBlue}{RGB}{28,49,117}

    \coordinate (top) at ($(pic cs:legend_top) + (-1mm,2.3mm)$);
    \coordinate (bottom) at ($(top) + (0,-83.75999999999999mm)$);

    \fill[GradientDarkBlue] (top) rectangle ($(top) + (6mm,-8.376mm)$);
    \shade[top color=GradientDarkBlue, bottom color=GradientCyan] ($(top) + (0,-8.376mm)$) rectangle ($(top) + (6mm,-16.752mm)$);
    \shade[top color=GradientCyan, bottom color=GradientTeal] ($(top) + (0,-16.752mm)$) rectangle ($(top) + (6mm,-25.127999999999997mm)$);
    \shade[top color=GradientTeal, bottom color=GradientGreen] ($(top) + (0,-25.127999999999997mm)$) rectangle ($(top) + (6mm,-33.504mm)$);
    \shade[top color=GradientGreen, bottom color=GradientLightGreen] ($(top) + (0,-33.504mm)$) rectangle ($(top) + (6mm,-41.879999999999995mm)$);
    \shade[top color=GradientLightGreen, bottom color=GradientYellow] ($(top) + (0,-41.879999999999995mm)$) rectangle ($(top) + (6mm,-50.25599999999999mm)$);
    \fill[GradientYellow] ($(top) + (0,-50.25599999999999mm)$) rectangle ($(bottom) + (6mm,0)$);

    \draw[black,thin] (top) rectangle ($(bottom) + (6mm,0)$);

    \draw[black] ($(top) + (6mm,0)$) -- ++(1mm,0) node[right,font=\scriptsize] {100\%};
    \draw[black] ($(top) + (6mm,-16.752mm)$) -- ++(1mm,0) node[right,font=\scriptsize] {80\%};
    \draw[black] ($(top) + (6mm,-33.504mm)$) -- ++(1mm,0) node[right,font=\scriptsize] {60\%};
    \draw[black] ($(top) + (6mm,-50.25599999999999mm)$) -- ++(1mm,0) node[right,font=\scriptsize] {40\%};
    \draw[black] ($(top) + (6mm,-67.008mm)$) -- ++(1mm,0) node[right,font=\scriptsize] {20\%};

    \node[above,font=\scriptsize\bfseries] at ($(top) + (5mm,6mm)$) {Agree \%};
\end{tikzpicture}
}
\label{tab:task_drivers}
\end{table*}

Next, we position tasks on the map and, by cluster, use qualitative insights to explain \textit{where} and \textit{why} developers seek/limit AI support.


\subsubsection{\textbf{Core work (C1)}\nopunct} comprised tasks central to \textit{development and quality-management}: coding, bug fixing, testing/QA, code review, system design, performance optimization, requirements engineering, security; as well as \textit{learning}, and \textit{research}. Appraised as high-value, high-stakes, and high-demand, most C1 tasks concentrated in \textbf{Build}/\textbf{Improve} (high-need) zones, indicating a strong appetite for AI support. However, identity alignment constrained delegation: participants sought AI as an \textit{augmentation} support while retaining ownership of core decisions, skills, and responsibilities. System design and requirements fell in \textbf{De-prioritize} due to AI's contextual misfit and reliability concerns.

\textsc{\textbf{Seek AI:}} For core work, participants used AI to boost \textbf{workflow efficiency}, delegating tedious steps: \pQuote{generate boilerplate code, build configurations, test cases...which I know how to write, but I don't want to write}{P353}; freeing focus for creative problem-solving: \pQuote{Leave me to do the fun [parts]}{P319}. They sought \textbf{proactive performance and quality assurance} to catch \pQuote{bugs, regressions, [performance] bottlenecks, and potential security issues early...where human review might miss patterns or edge cases}{P241}. This required \textbf{multi- and cross-context awareness}---AI that integrates signals across artefacts: \pQuote{It needs to look at logs, performance counters, etc., understand runtime behavior...then make changes, inspect results, try again}{P195}. P201 noted, \pQuote{We need AI to do a MUCH better job of analyzing a current code base/architecture so it can understand what/how to add/extend.
}{P201}. Participants envisioned \textbf{AI as a collaborator} aiding pair programming/debugging/testing/review; without overriding human judgment: \pQuote{The focus should be on AI making ME better at my job}{P213}.

For learning/research, participants wanted AI as a \textbf{personalized guide}, a \textbf{tool for information retrieval/synthesis}, and a \textbf{scaffold for ideation and practice}. They stressed adaptive help aligned with skills and goals: \pQuote{it should build my concepts from the ground up...based on my learning styles}{P281}. They valued AI's ability to surface relevant material: \pQuote{Researching could be where AI plays a big role...it can easily pick out relevant information from large amounts of data}{P28}, while cautioning against shortcuts where \pQuote{AI does all the work and [they] miss a chance to learn by doing}{P66}.

\textsc{\textbf{Limit AI:}} Participants resisted full automation of core work, insisting on \textbf{oversight and decision control}: \pQuote{I can't fully delegate the final code review to AI—my approval puts my name on it}{P117}. Others echoed, \pQuote{deciding whether to ship something with limitations or communicate a risk to leadership requires context, experience, and intuition that AI can’t fully grasp}{P301}. They emphasized that AI should not absolve human accountability: \pQuote{I wouldn’t want AI to handle final decision-making in high-stakes scenarios; responsibility should remain with experienced professionals}{P9}. 

They also \textbf{resisted AI to preserve professional identity/craft}: \pQuote{I do not want AI to handle writing code for me. That’s the part I enjoy and is the core of my work}{P110}. They warned that \textbf{overreliance risks deskilling}: \pQuote{intellectual offloading can result in errors that eventually no one understands}{P409}. P16 echoed: \pQuote{AI should not replace tasks that allow us to become better engineers}{P16}.

\textbf{Trust and quality concerns} reinforced these limits. Participants cited hallucinations, lack of transparency, and weak contextual support as reasons to keep AI in a supporting role: \pQuote{AI should not be the determining factor in how to solve quality problems...it frequently hallucinates with absolute certainty}{P17}. They were \textbf{wary of AI handling sensitive data}: \pQuote{I don't want AI to directly handle sensitive information, as I don't trust that information it sees stays truly secure}{P397}. Further, they flagged \textbf{AI-induced technical debt}: \pQuote{AI-generated code is often not readable or maintainable, which reduces long-term sustainability}{P149}.

System design \& requirements fell in \textbf{De-prioritize} due to AI's \textbf{capability gaps} and \textbf{contextual misfit}: \pQuote{These decisions require deep domain knowledge and long-term vision that only experienced engineers can provide}{P241}. Participants preferred human judgment, warning that \textbf{homogenized AI outputs} risk stifling innovation: \pQuote{AI's system design solutions bias toward old known solutions rather than a modern solution that solves the problem better}{P188}.

\begin{takeawayBox}

\textbf{Takeaway}: Developers seek AI as a collaborator in core work to boost efficiency and redirect focus on higher-order problem-solving (H1, H4), while retaining oversight and decision control in high-stakes, identity-laden aspects (H2, H3). 
\end{takeawayBox}

\subsubsection{\textbf{People \& AI-Building (C2)}\nopunct} covered AI integration (developing/embedding AI features) and mentoring. Appraisal-wise, C2 tasks showed strong identity alignment but moderate value, demands, and accountability, placing them in \textbf{Sustain}/\textbf{De-prioritize} zones (low AI \textit{need}). Participants preferred doing these tasks themselves, citing intrinsic motivations.

\textsc{\textbf{Limit AI:}} For AI integration, participants \textbf{resisted AI to preserve craft}: \pQuote{I don’t want AI to handle AI development, as it brings satisfaction to my work and requires craftsmanship}{P285}, and favored \textbf{deterministic workflows over stochastic outputs}: \pQuote{My team is focused on integrating AI into products. I don’t feel AI is up to any help for this...using it instead of a predictable workflow will not work out in the end}{P66}.

Mentoring drew stronger resistance given its relational nature: \pQuote{I don't want it to help mentor people. Relationships are important}{P85}. Participants viewed mentoring as \textbf{fundamentally interpersonal}---building trust, connections, and culture: \pQuote{new members need to interact with their team to build relationships. AI can't satisfactorily replace that for many, many years}{P122}. While AI might help with \pQuote{rote onboarding steps}{P70}, \textbf{AI misguidance can harm mentees}: \pQuote{The cost of it getting it wrong is terrible—and it will get it wrong}{P357}. 
Mentoring also serves as a \textbf{growth opportunity for mentors}: \pQuote{Mentoring teaches the mentor as well...humans need to do it to grow themselves}{P228}, reinforcing the need to be human-led.

\begin{takeawayBox}
\textbf{Takeaway}: For identity-laden and interpersonal work, developers resist AI to preserve craft, relationships, and personal growth (H2). AI is, at best, peripheral.
\end{takeawayBox}

\subsubsection{\textbf{Ops \& Coordination (C3)}\nopunct} covered (a) ops/maintenance toil (\textit{``run-the-systems''}): DevOps (CI/CD), environment setup, code maintenance, infrastructure monitoring, documentation; and (b) coordination/support (\textit{``relational work''}) overhead: project planning/management, stakeholder communications, customer support. Appraisals were moderate–high on \textit{value, demands, accountability} but low on \textit{identity}. Reported \textit{need} for AI was moderate, yet adoption lagged due to AI limitations, context fit, and trust concerns. In Fig.~\ref{fig:rq1-plots}, \textit{``run-the-systems''} tasks concentrate in \textbf{Build}, while ``relational-work'' sits in \textbf{De-prioritize} zone.

\textsc{\textbf{Seek AI:}} Participants wanted AI to reduce grunt C3 work (toil), provided it was reliable, deterministic, and context-aware.
For \textit{``run-the-systems''}, they sought an \textbf{assistant for well-scoped, effort-heavy tasks} (e.g., setup, maintenance, monitoring); those low in creative value. 
They also wanted AI to provision environments, automate upgrades/migrations, maintain system health, and update documentation. CI/CD was a recurring pain point: \pQuote{AI should help in maintenance of services, making sure that the lights are kept on when the developers move on to new features. Keep systems healthy; where safe, triage and remediate known issues}{P261}. 
Participants emphasized \textbf{cross-context awareness}---linking telemetry and artifacts to infer, triage, and predict failures: \pQuote{constantly analyze logs, metrics, and system behavior to identify deviations, triage causes, and predict potential failures before they impact operations or users}{PID309}. For large-scale refactoring (situated in the \textbf{Improve} zone), participants wanted architecture-aware changes applied safely across the codebase. For \textit{``relational work''}, AI was mostly used to \textbf{handle logistics}: drafting updates, summarizing meetings, pulling context, scheduling follow-ups—and for PM support (context-aware plans, dependency tracking), while strategy remained human-led.


\textbf{(Limit AI)} Participants largely \textbf{de-prioritized} AI for \textit{``relational work''}, noting that \textbf{contextual intuition, empathy, and long-term vision aren’t automatable}: \pQuote{I don’t want AI to handle ‘interpretation,’ drive product vision, or make strategic trade-off decisions}{PID3}. 
For stakeholder communications, they emphasized \textbf{authenticity \& relationship-building}: \pQuote{I don’t want AI to handle stakeholder comms...these require empathy, trust-building, and nuanced understanding of human dynamics}{PID172}; \pQuote{Client communication needs personal touch. Summarizing meetings is one thing, but replacing real touch points is too much}{PID217}. Customer support drew similar pushback: \pQuote{As a customer, being forced to an AI is frustrating}{P41}, with accuracy lapses risking support quality and brand damage. These concerns, alongside AI’s \textbf{capability gaps} (hallucinations, weak grounding, high prompt overhead), kept it backstage while \pQuote{humans hit send}{P47}.

By contrast, \textit{``run-the-systems''} tasks met less resistance in principle, but faced \textbf{trust and quality concerns} with (and/or absence of) current tools, that kept them in \textbf{Build} zone. Help was welcome only with determinism, verifiability, and human-gated control: \pQuote{Anything that touches prod stays behind human gates—no auto-deploys, no direct live changes, no publishing without supervision and transparency}{P165}.
Tool performance lapses made developers wary of AI refactoring at scale.—\pQuote{AI should not perform any large-scale refactoring}{PID182}. 
Finally, participants warned against \textbf{over-automation eroding operational intuition}: \pQuote{Engineers still need to learn how things work... AI should guide, not replace
}{PID16}.

\begin{takeawayBox}
\textbf{Takeaway}: Developers offload ops/coordination toil (H2, H4) only when AI is reliable, safe, and context-aware (H1, H3), resisting over-automation that erodes intuition or adds debt. Relational work aspects remain off-limits---empathy, intuition, and authenticity are irreducibly human.
\end{takeawayBox}

\subsection{RQ2: Which RAI design principles do developers prioritize in AI for SE tasks?}
\label{rq2}


Recall, participants selected five of eight RAI principles for relevant task categories (\textsection~\ref{sec:method}), reflecting priorities under forced trade-offs.

\begin{figure}[!hbt]
\centering
\includegraphics[width=\linewidth]{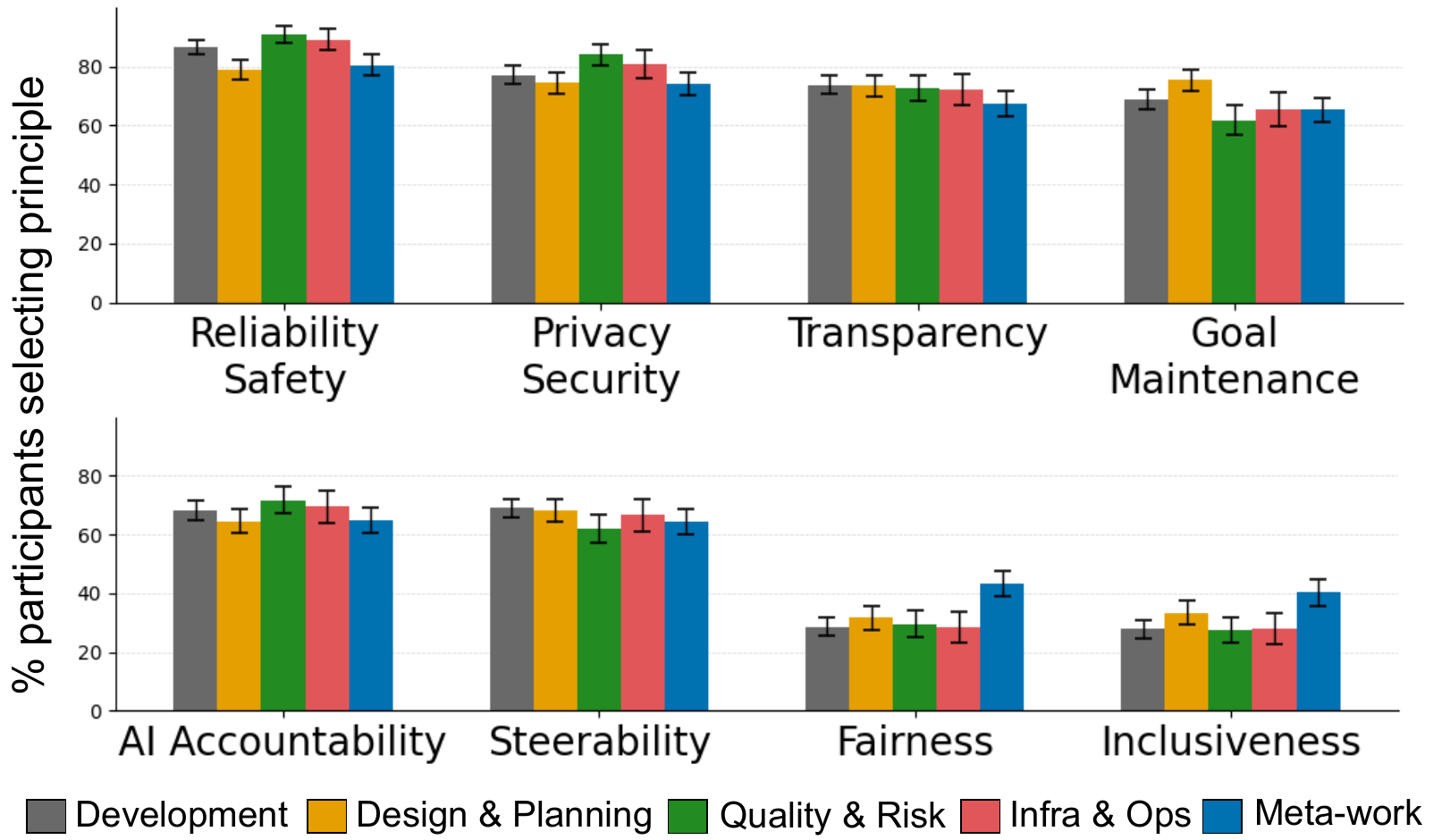}
\caption{Participants (\%) selecting each RAI principle as their top-5 priority for AI support across SE task categories; percentages reflect frequency and do not sum to 100\%.}
\label{fig:design_principle}
\end{figure}

As shown in Fig.~\ref{fig:design_principle}, participants most frequently selected \textbf{Reliability \& Safety} (85\%), \textbf{Privacy \& Security} (77\%), \textbf{Transparency} (72\%), \textbf{Goal Maintenance} (68\%), \textbf{AI Accountability} (67\%), and \textbf{Steerability} (67\%) across categories. \textbf{Fairness} (32\%) and \textbf{Inclusiveness} (32\%) followed. Overall, they prioritized pre-requisites ensuring correctness, security, alignment, and steerability before expecting credible support for broader humanist principles: \pQuote{Surely all of them are important but at which stage? Right now, the basics aren't even done well, so those are [what] I selected}{P43}.


To examine how RAI priorities varied, we fit logistic Generalized Linear Mixed Models (GLMMs)~\cite{breslow1993approximate} per principle (Tab.~\ref{rq2-priorities}). The models predicted whether a principle was prioritized ($Outcome=0/1$) as a function of task categories, mean-centered SE \& AI experience, risk tolerance, and technophilia, with random effects for within-\textit{participant} repeated measures. The intercept (baseline) represents a developer with average SE, AI experience, and AI-dispositions (relative to their peers) prioritizing a principle in AI support for \textit{development-heavy work}. Tab.~\ref{rq2-priorities} reports odds ratios (ORs) relative to this baseline (OR$>$1 = higher odds and vice versa). For example, baseline odds of prioritizing \textit{Privacy \& Security} in AI for development tasks were 8.19 (i.e., $8.19/(1+8.19)$ $\approx$89\% probability); for quality/risk-management, odds increased by 1.91 ($8.19 \times 1.91 = 15.65$, $\approx$94\% probability). 

Next, we integrate qualitative insights to explain why these patterns emerged and how priorities shifted across SE work.

\textbf{Interpretation guideline}: These results reflect developers' RAI priorities under a forced-choice design; they neither prescribe policy nor imply that any RAI principle is optional. \textbf{All remain essential}. Read them as a pragmatic "order of operations" for current tools, varying by task category and individual dispositions.

\subsubsection{\textbf{How do priorities vary across task categories?}} \hfill

\textbf{(1)} For \textbf{\textit{systems-facing work}} (development, infrastructure/ops, quality \& risk)---concentrated in the \textbf{Build/Improve} zones of Fig.~\ref{fig:rq1-plots}---participants imposed strict gating for trusting AI. \textbf{Reliability \& Safety} (base odds: 18.15; $\approx$95\%) and \textbf{Privacy \& Security} (base odds: 8.19; $\approx$89\%) were non-negotiable. They stressed that AI errors did not \textit{“net to zero”}---they wasted time, sent teams down unproductive paths, and amplified risk: \pQuote{AI MUST BE correct for it to be useful. Incorrect AI may as well be throwing spaghetti at a wall—it’s more work to fix it}{P83}. 
Privacy concerns intensified in infra/ops (OR = 1.38) and quality/risk (OR = 1.91), where artifacts were sensitive: \pQuote{One privacy slip and trust collapses}{P313}. Respondents demanded \pQuote{absolute assurance that AI safeguards information, preventing leaks or unauthorized access}{P322} and declined tools \pQuote{for security work unless [they’re] secure...and sources [are] clear}{P435}.

Participants next emphasized \textbf{Transparency} (base odds: 5.17; $\approx$84\%) to verify assumptions, catch hallucinations, and justify (or revise) AI contributions. They tied this need to personal accountability and learning: \pQuote{I feel accountable for my work, so I feel accountable for what the AI has done. I want it to explain why an action was taken. This also helps me grow as a developer}{P36}.

\begin{table*}[!ht]
\footnotesize
\centering
\caption{\small Logistic Generalized Linear Mixed-effects Models (GLMMs) predicting baseline odds and odds ratios (ORs) of Responsible AI design priorities by task category, developer experience, and AI dispositions. Experience and disposition predictors are mean-centered. The constant (baseline) reflects the odds of a developer performing development-heavy tasks with average SE experience, AI experience, risk tolerance, and technophilic motivations. Odds ratios are relative to the baseline odds; priorities differ from baseline only if statistically significant.}
\label{tab:regression-haiaspects}
\begin{tabular}{l*{8}{r}}
\toprule
\toprule
& \multicolumn{8}{c}{\textit{Dependent variable}} \\
\cmidrule(lr){2-9}
\textbf{Factor} 
    & \makecell{Reliability \&\\Safety} 
    & \makecell{Privacy \& \\Security} 
    & \makecell{Transparency} 
    & \makecell{Goal\\Maintenance} 
    & Steerability
    & \makecell{AI\\Accountability} 
    & Fairness
    & Inclusiveness\\
\midrule
\textbf{Constant} - Base (Development) odds & 18.15*** & 8.19*** & 5.17*** & 4.68*** & 3.65*** & 3.03*** & 0.21*** & 0.2*** \\
\midrule
\multicolumn{9}{l}{\textbf{Task Categories}} \\
Design \& Planning & 0.49** &  &  & 1.45* &  &  & 1.48* & 1.61** \\
Quality \& Risk Management & & 1.91** &  & 0.61** &  &  & 1.07** & \\
Infrastructure \& Operations & & 1.38* &  &  & & &  & \\
Meta-work                            & &  &  &  &  &  & 3.06*** & 2.49*** \\

\midrule

\multicolumn{9}{l}{\textbf{Experience \& Dispositions (mean-centered)}} \\
SE Experience  & 1.15* & & & & 1.21* & & &  \\
AI Experience  & & & 1.3* &  & 1.11* & & & \\

Risk Tolerance  & & & & & 1.13** &  & & \\
Technophilic Motivations & & & & 1.16** & 1.28* & & & \\
\midrule

$R^2_{\mathrm{m}} / R^2_{\mathrm{c}}$ & 0.06 / 0.32 & 0.03 / 0.31 & 0.05 / 0.29  & 0.07 / 0.31  & 0.04 / 0.37 & 0.04 / 0.43 & 0.07 / 0.44 & 0.03 / 0.43      \\
\bottomrule

\end{tabular}
\begin{tablenotes}
\item \textit{Note: p-values are adjusted for False Discovery Rate (FDR), using Benjamini-Hochberg \cite{thissen2002quick}. Blank cells indicate odds equal to the baseline. $^{*}p<.05$; $^{**}p<.01$; $^{***}p<.001$}
\end{tablenotes}
\label{rq2-priorities}
\end{table*}

To keep tools aligned with shifting objectives, they highlighted \textbf{Goal Maintenance} (base odds: 4.68; $\approx$82\%): \pQuote{The things I need to be giving my attention to are changing all the time, so if AI could keep up with that via Goal Maintenance, that would be huge}{P791}. To address drift, participants prioritized \textbf{Steerability} (base odds: 3.65; $\approx$79\%) for redirecting AI and \textbf{AI Accountability} (base odds: 3.03; $\approx$75\%) for provenance to backtrack and trace errors: \pQuote{Lack of steerability \& accountability makes it hard to use AI for tasks that require extensive time/detail...it’s hard to put it back on track and/or tell where it went wrong...I often have to start new chats, which is frustrating, losing progress because the context is gone}{P495}.

\textbf{Fairness} surfaced as contextually salient (OR = 1.07) in quality/risk management; since reviews impacted releases and defect attribution. Participants stressed the need for unbiased evaluation: \pQuote{Fairness of PR review is hard for humans, so I want it to be a success metric for AI reviewers}{P196}; \pQuote{It is important to have unbiased AI for quality-related tasks to ensure the integrity of the work is not compromised}{P461}. 
Some saw fairness as nested within reliability and safety; others noted a tension between fairness and privacy, as bias checks might require \pQuote{exposing personal attributes the AI shouldn’t know}{P476}.

\textbf{(2)} For \textbf{\textit{design and human-facing work}} (design/planning, meta-work)---concentrated in the \textbf{De-prioritize} zone of Fig.~\ref{fig:rq1-plots}---participants elevated \textbf{Fairness} (OR = 1.48, 3.06) and \textbf{Inclusiveness} (OR = 1.61, 2.49). They wanted AI to broaden perspectives and avoid reinforcing bias in collaborative or client-facing contexts: \pQuote{Inclusiveness and fairness of AI features should be baked into design planning from the start...it needs a grasp of business requirements and diversity of audiences}{P653}. This was especially salient for documentation and stakeholder communications: \pQuote{AI must ensure inclusiveness and fairness so content works for all customers}{P120}.

Other priorities were consistent with the baseline (e.g., Transparency for learning/research, Privacy for sensitive communications, Steerability, and AI Accountability for system design). In design/planning, however, participants downweighted \textbf{Reliability} when AI served as an ideation scaffold (OR = 0.49), and prioritized \textbf{Goal Maintenance} (OR = 1.45). 
When AI scaffolded creativity, adaptability was valued: \pQuote{Creativity of AI is important; I'm willing to tolerate errors}{P180}. Participants appreciated AI surfacing options that sparked innovation (even if imperfect), provided it stayed aligned with (evolving) objectives: \pQuote{During planning ... AI needs to keep up with evolution. I'd also expect AI to bring in much more outside perspectives to synthesize a range of feedback}{P459}.

\begin{takeawayBox}
    \textbf{Takeaway}: In systems-facing work, \textbf{Reliability} and \textbf{Privacy} are central; then comes \textbf{Transparency}, \textbf{Goal Maintenance}, \textbf{Steerability}, and \textbf{AI Accountability}. \textbf{Fairness} and \textbf{Inclusiveness} are elevated in design and human-facing work. Developers relax \textbf{Reliability} for creativity and emphasize \textbf{Goal Maintenance} as needs shift.
    
\end{takeawayBox}

\subsubsection{\textbf{How do priorities vary by experience/AI dispositions?}} \hfill

\textbf{Steerability} rose in priority with higher \textit{SE experience} (OR = 1.21), \textit{AI experience} (OR = 1.11), \textit{risk tolerance} (OR = 1.13), and \textit{technophilia} (OR = 1.28). Viewed through Self-Determination Theory~\cite{deci2000and, ryan2000self}, this pattern reflected protection of \textit{autonomy} and \textit{competence}: experienced developers favored control that kept AI actions interruptible and easy to correct, citing course-correction overheads---\pQuote{Sometimes it spirals off... backtracking is harder than starting over}{P657}. 
Risk-tolerant individuals treated steerability as a safeguard for rapid intervention. Participants also resisted modes (e.g., bulk edits) that could erode competence over time: \pQuote{Multi-file edit modes feel to take away steerability...
I've noticed it harms the engineer's skills over the long term more than it helps}{P754}. 

Experienced SEs prioritized \textbf{Reliability \& Safety} (OR = 1.15), consistent with their sharper awareness of downstream \textit{``automation surprises''}~\cite{parasuraman2010complacency}. Those with more AI experience prioritized \textbf{Transparency} (OR = 1.30), as familiarity with AI’s quirks heightened demands for verification and provenance: \pQuote{debug and justify outputs even when they look plausible}{P367}.

Technophilic individuals prioritized \textbf{Goal Maintenance} (OR = 1.16). As recent work notes~\cite{choudhuri2025needs}, their intrinsic drive to explore AI tooling collides with current frictions (prompt churn, drift, and limited affordances), raising the cognitive \textit{cost of exploration}, which ultimately erodes engagement~\cite{deci2000and, unsworth2012variation}.

\vspace{2mm}
\begin{takeawayBox}
\textbf{Takeaway}: Individual traits shape RAI priorities: SE experience heightens demands for \textbf{Reliability \& Safety}, AI experience for \textbf{Transparency}, and technophilia for \textbf{Goal Maintenance}. SE/AI experience, risk tolerance, and technophilia all amplify emphasis on \textbf{Steerability}, reflecting a strong need for agency.

\end{takeawayBox}

\section{Discussion}
\label{sec:discussion}

Our results capture a snapshot of the evolving developer-AI collaboration landscape. Findings may shift as tools evolve, yet both current patterns and their theoretical grounding are informative. Task forms and placements may change as capabilities grow, yet deeper appraisals (e.g., enduring needs in quality and people work) are likely to persist. 
Our mixed-methods clustering approach can relocate tasks as the landscape evolves. Current frictions in reliability, security, and transparency point to priority investments, especially in the ``outer loop'' (e.g., testing, review, release, and governance).

\subsection{Implications for practice}
\label{implications-practice}
\textit{\textbf{Augmentation over blunt automation}:} Developers prefer AI that amplifies creativity and craft, not just removes toil, consistent with evidence linking meaningful work to growth and contribution~\cite{lips2022we}. The \textit{``human / AI / human+AI''} framing~\cite{msft_nfw} applies: some activities are best human-led, some AI-led, and many human+AI. Our map shows where each mode fits today and where to invest to enhance rather than hollow out meaningful work.

Developers want AI as a cognitive collaborator on core work---helping decompose problems, generate alternatives, and pivot across artifacts---while preserving agency and craft. Design for:

\textbf{(1)} \textit{Provenance and transparency}: show sources, explanations, confidence, and transformations; keep decision paths
inspectable; maintain traceable links among artifacts.

\textbf{(2)} \textit{Decision control}: default to suggest-only flows with reversible changes, batched diffs with rationales, and explicit approval checks.

\textbf{(3)} \textit{Craft-preserving AI design}: reveal intermediate reasoning so developers learn, avoiding skill erosion from over-automation.

Where work depends on connection and recognition, developers de-prioritize AI. The right stance is \textit{peripheral support}: assist with preparation (briefings, what-if scripts), reflection (summaries, action extraction), and equity (bias checks, inclusivity), while leaving human contact and credit intact. This \textit{``complement, don’t crowd out''} principle mitigates AI intrusion into social labor.

Automation often \emph{shifts} toil rather than eliminating it~\cite{acemoglu2019automation, carver1924elements}. Time saved can reappear in setup, oversight, or remediation. Highest returns pair automation with reliability, transparency, and alignment: strong grounding; guardrails for hallucinations or unsafe edits; test-first and integration-aware suggestions that respect CI, policy, and compliance. Human oversight remains essential as software work is inherently socio-technical and consequential.

In practice: ship for complementarity in \emph{Core Work}, treat \emph{Ops \& Coordination} as a reliability problem first, and keep \emph{People \& AI-Building} human-led with AI assistive.





\textit{\textbf{Deliberate job (re)design}:} Teams can use the map to shift time from toil to higher-order knowledge and people work, creating room for learning and designing ceremonies that preserve recognition (e.g., crediting rationales and reviews). Leaders should track experience outcomes (flow, satisfaction, confidence) alongside throughput and invest in intentional \textit{``moments that matter''} to maintain cohesion in an era of AI-accelerated solo work. This shift is not automatic; it requires intentional \emph{job crafting} (redesigning roles, rituals, recognition) so higher-order work is visible/rewarded; and support for \emph{horizontal skill expansion} (product sense, data/AI literacy, operations). Open questions remain: \textit{Does AI truly create time for meaningful work, or mainly boost throughput? Which job-crafting moves best preserve meaningful work? Under what conditions?}

\vspace{-2mm}
\subsection{Implications for research}
\label{implications-research}

As assistants grow more agentic, three priorities emerge:

\textbf{(1) Transparency \& observability.} What forms of evidence (e.g., decision logs, rationales) improve oversight without inducing overreliance~\cite{buccinca2021trust}? \emph{Needed:} validated measures of “useful transparency” and experiments on trust calibration/error detection.

\textbf{(2) Goal maintenance.} How should evolving goals and constraints be represented so agents detect and prevent drift across artifacts/sessions? \emph{Needed:} shared human- \& AI-legible goal schemas; drift benchmarks; causal tests of guardrails (pre-commit checks, test-first prompts) on quality, latency, and friction.

\textbf{(3) Steerability \& developer agency.} Which interrupt/redirect mechanisms and delegation policies best balance control under varying risk? \emph{Needed:} task-typed autonomy ladders/taxonomies; evaluations tying agency to outcome quality.


We present our approach (cognitive appraisal + mixed-methods clustering) as a \emph{living instrument}. Periodic re-runs can relocate tasks, recalibrate RAI priorities, and test whether tooling improvements measurably shift developer experiences and outcomes.

\subsection{Limitations}
\label{sec:threat}

\textbf{\textit{Construct validity}}
We measured constructs using self-reported items grounded in established theory. Still, surveys can introduce bias or misunderstanding. We guarded against this risk by involving practitioners in design, sandboxing and piloting, randomizing blocks, adding attention checks, and screening patterned responses. 

\textbf{\textit{Internal validity}}
As a cross-sectional study~\cite{stol2018abc}, we report associations, not causation. Self-selection bias is possible, since those with stronger views may be more likely to participate. We strengthened validity by triangulating quantitative results with coded qualitative data and theory, and using member checking. As with all survey-based work, results reflect self-reported perceptions. 
Interpret RAI prioritization with care, since a normative ``ought'' does not follow from an empirical ``is''~\cite{musschenga2005empirical}. Our goal is to inform context-sensitive RAI choices in SE and offer critical reflections, not to prescribe one course of action. Following prior recommendations~\cite{denzin2011sage}, we do not report frequencies for qualitative findings.

\textbf{\textit{External validity}}
We studied Microsoft developers across global sites, diverse teams and roles, many domains, and varied processes. This scope 
may not generalize to smaller organizations or OSS. We do not claim to represent all software engineers. Instead, we provide an in-depth account of a large organizational context. Single case studies have advanced scientific discovery~\cite{flyvbjerg2006five} and produced insights in social science and software engineering~\cite{storey2019towards,kuper2004social}. 
Future work should test transferability in other contexts.
\section{Conclusion}
\label{sec:conclusion}



AI in SE should complement, not replace, developers. Demand is highest for tools that improve core work and cut toil, with clear limits around identity-centric tasks. Developers prioritize responsible AI: reliable, safe, privacy-preserving, transparent, and steerable, so they can stay in control and learn. Design goal-aware, observable, interruptible systems, and invest in AI where it matters.

\textbf{Data availability.} Supplementary materials are in~\cite{supplemental}; an interactive report is at
\textcolor{blue}{\url{https://aka.ms/AI-Where-It-Matters}}.

\begin{acks}
We thank Emerson Murphy Hill, Sian Lindley, and Saleema Amershi for their valuable feedback. We also thank all the survey respondents for their time and insights. Rudrajit Choudhuri completed this work during a summer internship at Microsoft Research.
\end{acks}


\bibliographystyle{ACM-Reference-Format}
\bibliography{bibfile}

\end{document}